\documentclass[12pt,preprint]{emulateapj}
\usepackage{apjfonts,psfig}

\shorttitle{SB profiles of LMC, SMC and Fornax GCs}
\shortauthors{Noyola and Gebhardt}

\begin{document}

\title{Surface Brightness Profiles for a sample of LMC, SMC and Fornax galaxy Globular Clusters}

\author{E. Noyola}
\affil{Astronomy Department, University of Texas at Austin, TX 78712}
\affil{Max-Planck-Institut f\"{u}r extraterrestrische Physik, Giessenbachstraße
 85748, Garching }

\email{noyola@mpe.mpg.de,gebhardt@astro.as.utexas.edu }

\author{K. Gebhardt}
\affil{Astronomy Department, University of Texas at Austin, TX 78712}

\begin{abstract}

We use Hubble Space Telescope archival images to measure central
surface brightness profiles of globular clusters around satellite
galaxies of the Milky Way. We report results for 21 clusters around
the LMC, 5 around the SMC, and 4 around the Fornax dwarf galaxy. The
profiles are obtained using a recently developed technique based on
measuring integrated light, which is tested on an extensive simulated
dataset. Our results show that for 70\% of the sample, the central
photometric points of our profiles are brighter than previous
measurements using star counts with deviations as large as
2~mag/acrsec$^2$. About 40\% of the objects have central profiles
deviating from a flat central core, with central logarithmic slopes
continuously distributed between -0.2 and -1.2. These results are
compared with those found for a sample of Galactic clusters using the
same method. We confirm the known correlation in which younger
clusters tend to have smaller core radii, and we find that they also
have brighter central surface brightness values. This seems to
indicate that globular clusters might be born relatively concentrated,
and that a profile with extended flat cores might not be the ideal
choice for initial profiles in theoretical models.
 
\end{abstract}

\keywords{globular clusters: general, stellar dynamics}

\section{Introduction}\label{intro}

The observational study of internal dynamics of globular clusters
(GCs) has benefited from imaging from space as well as enhanced
spectroscopic capabilities in the ground. \citet{noy06} (From now on
called ``Paper~I'') measure surface brightness profiles (SB) from
$\it{Hubble~Space~Telescope}$ ($\it{HST}$) images for a sample of 38
galactic globular clusters. The results from that work show that half
of the objects in the sample are not consistent with having central
flat cores, but instead, the distribution central surface brightness
logarithmic slopes is continuous form $-0.2$ to $-0.8$. The ages of
the Galactic clusters are all confined to a narrow range older than
$\sim$10Gyr \citep{sal02,ang05}. It is desirable to measure central SB
profiles of globular clusters with younger populations to find out if
these central cusps are also observed in less evolved
clusters. Globular clusters around Milky Way satellites are ideal
targets for this task since they have a larger age range, they are
relatively near, and many of them have been observed with $\it{HST}$.

Surface brightness profiles have been obtained for GCs in the Large
Magellanic Cloud (LMC), Small Magellanic Cloud (SMC), and Fornax dwarf
galaxies in various studies using ground-based data. For the LMC
clusters, star counts \citep{kon87a}, aperture photometry
\citep{mat87,els91} and hybrid techniques \citep{els87} have been used
to obtain surface density profiles for a variety of subgroups (rich,
old, young, disk, and halo clusters). For the SMC clusters, only a few
studies have measured density profiles from star counts
\citep{kon83,kon86}. A couple of studies measure density profiles
from aperture photometry for globular clusters around the Fornax dwarf
galaxy \citep{smi96,rod94}. All of these studies are very useful for
studying SB profiles at large radius, but at small radius they suffer
from the usual seeing and crowding problems associated with
ground-based observations.

A large systematic study of surface brightness profiles obtained from
space-based imaging has been carried out by Mackey \& Gilmore
(2003a,b,c) (from now on collectively referred to as MAC03). They
gather a broad sample of LMC, SMC and Fornax galaxy GCs imaged with
WFPC2. They obtain SB profiles by measuring star counts weighted by
brightness from which they derive fundamental quantities like central
density and core radius by fitting EFF profiles \citep{els87,els91},
which are power-law plus core profiles with three parameters: core
radius, central surface brightness, and slope of the power-law. They
determine that 20$\pm$7\% of the clusters in their sample are
consistent with a post-core-collapse morphology, a similar number to
the one found for Galactic clusters \citep{tra95}. When they compare
their profiles with previous results obtained from ground based
images, they find that important aspects of the nature of the profiles
can be measured by improving the spatial resolution. \citet{mcl05}
(from now on called MVM05) combine the data from MAC03 with star
counts profiles from ground-based data in order to obtain a more
accurate photometric normalization. They fit the re-normalized
de-reddened resulting profiles with variety of models such as King
fits \citep{kin66}, an alternate modified isothermal model by
\citet{wil75}, which has more extended envelopes than a King model,
and a power-law plus core model like the one used in MAC03. They
conclude that The Wilson fits provide the best description of the
outer part of the clusters for both old and young populations.

\citet{els89} and \citet{els92} find an interesting relation between
core radii and age for a sample of LMC globular clusters in which the
core radius seems to increase with ages between 1 Myr and 1 Gyr and
then shows a wide range of values for older clusters. Using $\it{HST}$
data, \citet{deg02} explore the matter for a sample of rich LMC
globular clusters and find that young clusters tend to have small core
radii while older clusters have an increasingly large spread of core
radii. MAC03 explored this relationship and found that the relation is
also valid for globular clusters around other Milky Way satellites
besides those in the LMC.

We concentrate in the central parts of the clusters since this is the
region for which our technique has found differences in the SB shape
when compared to profiles obtained from star counts for some
clusters. Improving the measurements in this region and merging the
results from our Galactic sample with those of this new sample, can
help to understand their dynamical evolution. The LMC, SMC and Fornax
galaxy globular cluster systems offer a unique window of opportunity
to test if there are fundamental differences between systems due to
their age.

\section{Simulations}\label{sim}

In paper~I we performed a large number of simulations in order to
establish the best method for measuring surface brightness profiles
from $\it{HST}$ images and also to estimate the uncertainties of our
measurements. Results from that paper indicate that the only way to
measure reliable surface brightness profiles from integrated light is
by using high signal to noise images. In order to evaluate how our
findings for galactic GCs translate to clusters further away, we again
perform extensive simulations, which we describe in detail in this
section.

\begin{figure}[t]
\centerline{\psfig{file=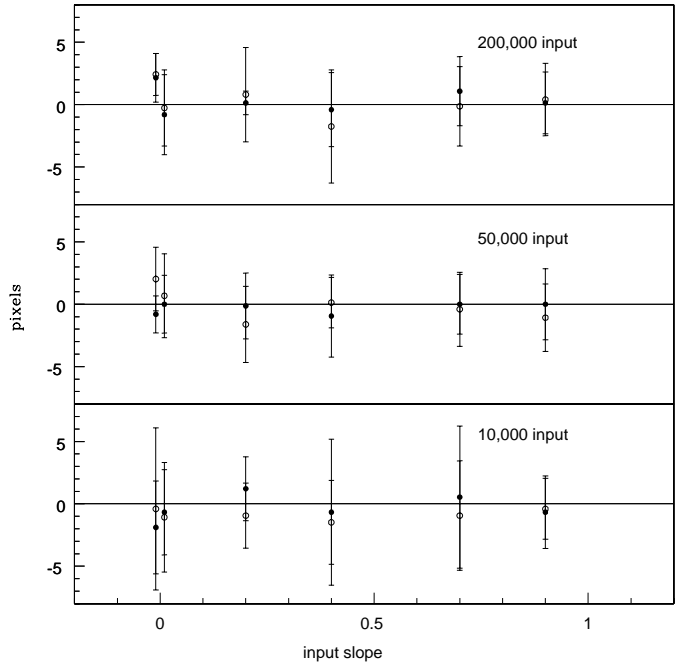,width=9.5cm,angle=0}}
\figcaption[Center measurement results for the simulated
  datasets]{Comparison between the measured and input center for every
  set of simulations with different central slopes. The average
  distance between the actual and measured center is shown for the $x$
  (solid points) and $y$ (open points) coordinates in pixels. A small
  horizontal offset is introduced for clarity for the two cases with
  central zero slope. Error bars are the standard deviation of the
  individual measurements for each case. Each PC pixel is
  0.046\arcsec.}
\label{ch2f1}
\end{figure}

\begin{deluxetable}{lccccc}
\tablewidth{0pt}
\tabletypesize{\Large}
\tablecaption{\label{tbl1b}Simulation input.}
\tablehead{
\colhead{model} &
\colhead{inner slope}   &
\colhead{outer slope} &
\colhead{break radius} &
\colhead{hardness of brake} &\\
\colhead{} &
\colhead{$\gamma$}   &
\colhead{$\beta$} &
\colhead{(pixels)} &
\colhead{$\alpha$} &
}
\startdata
model 1 & 0.0 & 1.8 & 85 & 2  \\
model 2 & 0.0 & 2.5 & 340 & 3 \\
model 3 & 0.2 & 2.5 & 90 & 1\\
model 4 & 0.4 & 1.6 & 90 & 2 \\
model 5 & 0.7 & 1.8 & 90 & 2 \\
model 6 & 0.9 & 2.0 & 90 & 1  \\
\enddata
\end{deluxetable}

\vspace{5pt}

\subsection{Image Construction}\label{sima}

The way we create a simulated image is by adding synthetic stars on a
background image using the task ADDSTAR in DAOPHOT \citep{ste87}. The
background image we use is a WFPC2 image of a very unpopulated field
for which the few present stars have been cleanly subtracted. The
input star lists are created in the same way as in paper~I. With a
given SB profile and a luminosity function, stars are generated
randomly around a given center (the middle of the chip) following the
two probability distributions, the surface brightness for radial
distribution, and the luminosity function for the magnitude
distribution. The supplied luminosity function comes from
\citet{jim98} and it is corrected using the distance modulus for the
LMC. The observed luminosity function is extended in the faint end in
order to include unresolved background light in the simulated images,
so we expect to recover fewer stars than the number we input.

\begin{figure*}[t]
\centerline{\psfig{file=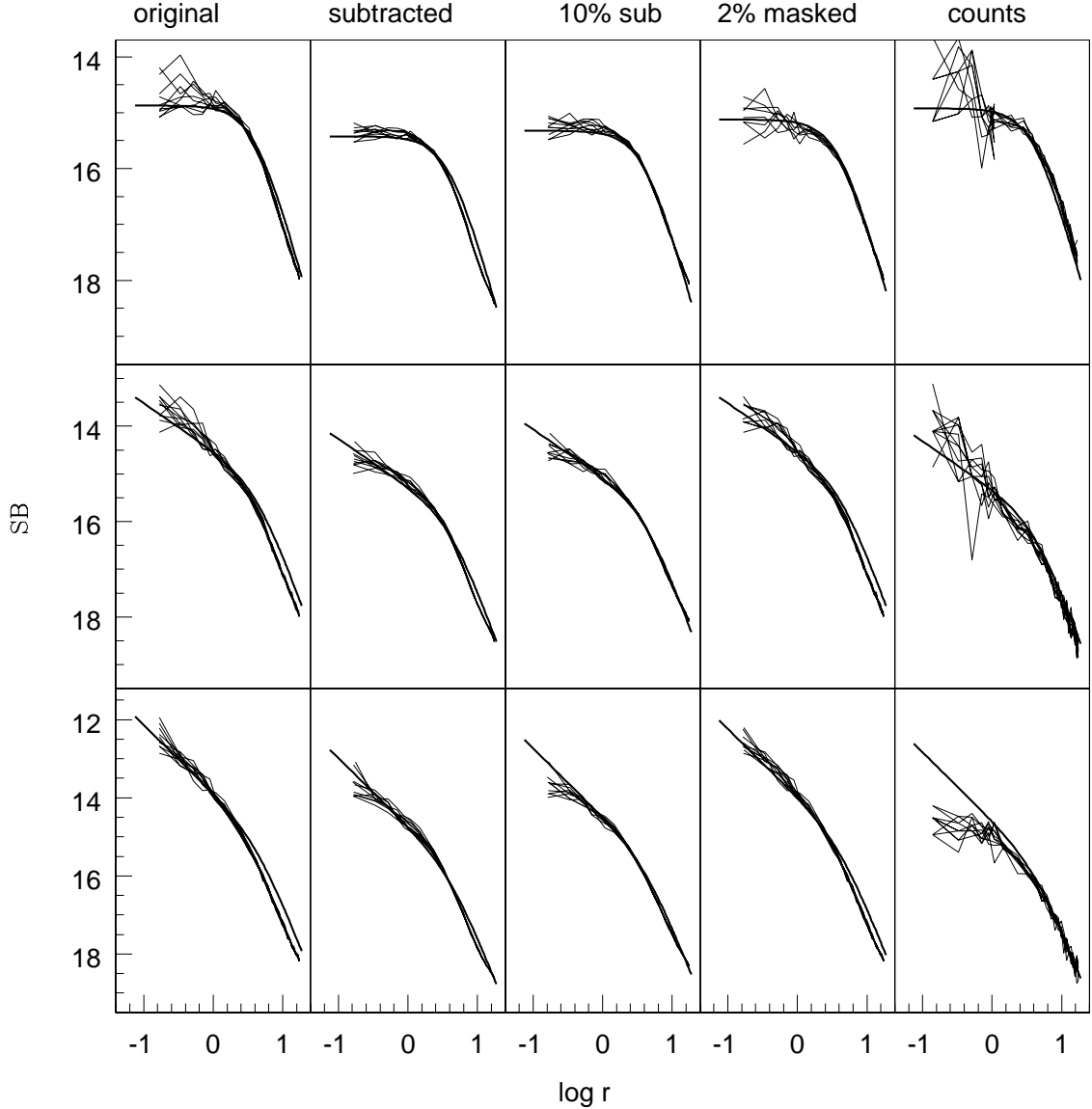,width=16.0cm,angle=0}}
\figcaption[Surface brightness results for simulated dataset]{Surface
  brightness profiles for three groups of simulations with 200,000
  input stars. For each case (models 1, 4, and 5) the measurements
  from individual realizations (thin lines) are plotted against the
  input profile (thick solid line). The profiles are measured from
  four different images: full, subtracted, 10\% brightest stars
  subtracted and 3\% brightest stars masked, and also from star
  counts. The vertical axis is on an arbitrary magnitude scale.}
\label{ch2f2}
\end{figure*}

\begin{figure*}[t]
\centerline{\psfig{file=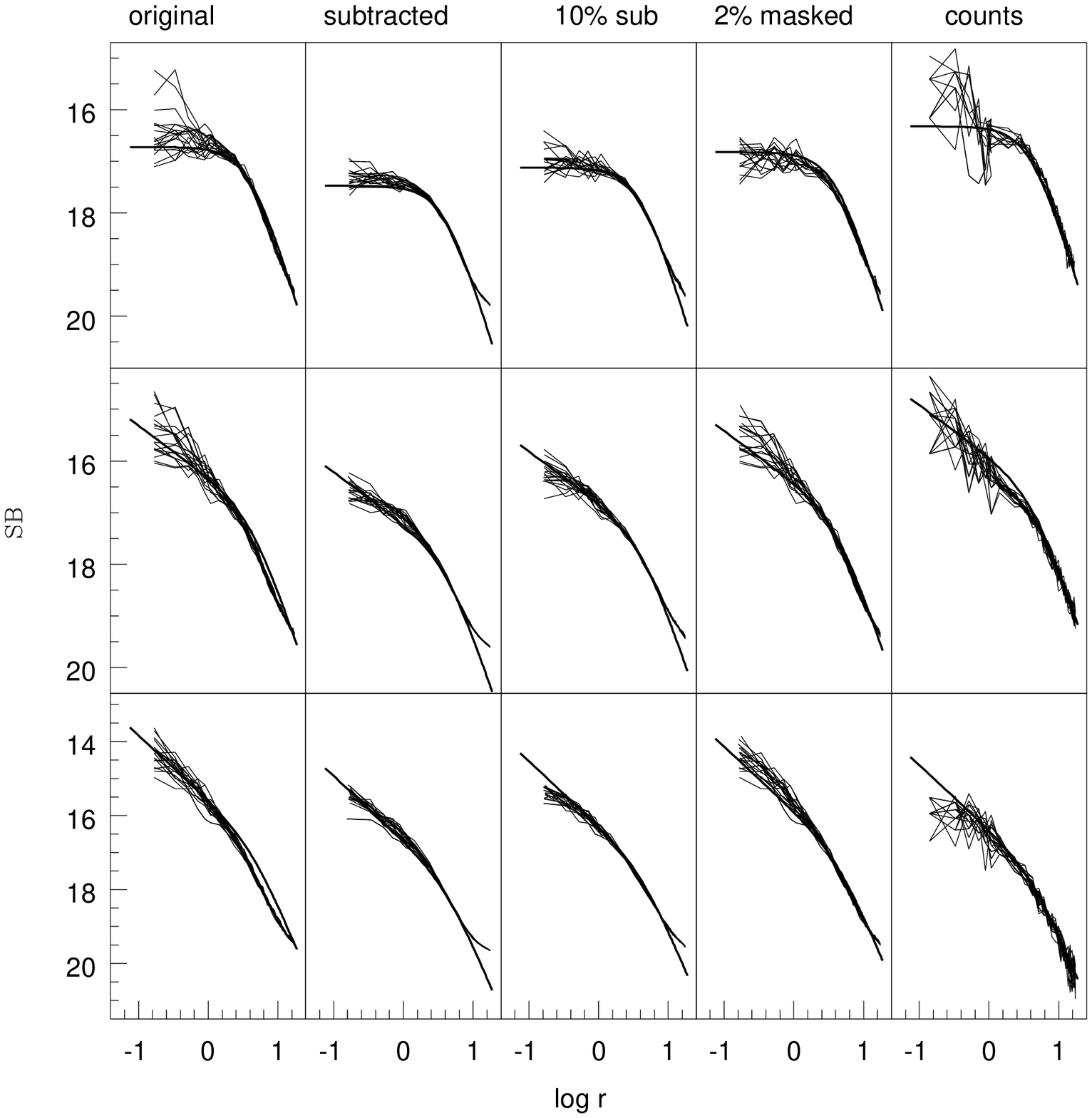,width=16.0cm,angle=0}}
\figcaption[More SB results for simulated datasets]{Same as previous
  figure but for simulations with 50,000 input stars.}
\label{ch2f3}
\end{figure*}

In paper~I, we simulate SB profiles with the shape of various power
laws. This gave us a good feel for our ability to recover a given
central slope, but we could not test our ability to measure turnover
radius. In order to better test our method, this time we create a
series of profiles formed by two power-laws joined at a break radius
with a variable sharpness of break, better known as Nuker profiles
\citep{lau95}. A Nuker profile is defined in the following way

$$ I(r)=I_b 2^\frac{(\beta-\gamma)}{\alpha}
\left({\frac{r}{r_b}}\right)^{-\gamma}
\left(1+\left(\frac{r}{r_b}\right)^\alpha\right)^\frac{(\gamma-\beta)}{\alpha}
,
$$

\vspace{5pt}

\noindent where $r_b$ is the break radius, $I_b$ is the surface
brightness at the break radius, $-\gamma$ is the asymptotic inner
slope, $-\beta$ is the asymptotic outer slope, and $\alpha$ is the
sharpness of break. By using these type of profiles, we are capable of
reproducing the characteristics of observed central profiles for the
sample. We create six different input profiles, whose parameters are
summarized in Table \ref{tbl1b}. The radial extend of the simulated
clusters is 400 pixels, which is equivalent to 18.4\arcsec\ with the
PC pixel scale (0.046 arcsec/pixel).

Once we have the input profiles, we proceed to create multiple
realizations of a given model including different numbers of
stars. Using various DAOPHOT tasks we add synthetic stars onto the
background image. We use as the input point spread function (PSF) the
one calculated for the LMC cluster NGC~1835 with a PSF radius of 9
pixels. Judging by the number of recovered stars in the real data, we
create images with three different amounts of input stars: 200,000
input stars, which yields $\sim$10,000 detected stars; 50,000 input
stars, giving $\sim$ 6,000 detected stars; and 10,000 input stars, for
which we find $\sim$ 2,000 stars. The majority of the real clusters in
the sample are comparable to the first two cases. The different
realizations have the exact same input parameters but come from
different, non-overlapping star lists. We create 10 realizations for
the 200,000 input stars case, and 20 for the other two. It is worth
noting that the number of detected stars decreases with increasing
input central slope for the same number of input stars. For the
steepest central slope $\sim$8,000 stars are found compared to the
$\sim$10,000 for the zero central slope cases. To avoid confusion, we
always refer to the simulated datasets by the number of input stars
rather than the number of detected stars.

\subsection{Center Determination}\label{scen}

Having an accurate estimate of the center position is a key step to
measuring an accurate density profile. Our technique for finding the
center of a cluster is described in detail on paper~I. We take a guess
center, divide the image in eight sectors converging at that center,
count the stars in each sector and calculate the standard deviation of
those eight numbers. We change to a different guess center and perform
the same operation. In the end we have a grid of guess centers with a
standard deviation value associated to them. We fit a surface using a
spline smoothing technique \citep{wab80,bat86} and choose the minimum
of this surface as our center.

We test the accuracy of our center determination technique by applying
it to these simulated images. Figure~\ref{ch2f1} shows the average
measured center and the standard deviation of the measurements for
different groups of simulations. The maximum deviation observed is of
$\sim$7 pixels, which is equivalent to $\sim$0.3\arcsec. These results
are better than those in paper~I. We believe the reason for this is
that there are more stars enclosed in the same projected radius due to
the distance difference, therefore the center estimation is improved.

\subsection{Surface Brightness Profiles}\label{ssb}

We compare the results of measuring the density profile from
integrated light versus doing it using star counts. We refer the
reader to the detailed discussion in section 2.3 of paper~I about the
strengths and weaknesses of each method. Results from that paper
indicate that using a robust estimator to calculate the number of
counts per pixel in a given area is the best way to recover the
central part of the profile. For that reason we use the same robust
estimator, the biweight \citep{bee90}, for our measurements in this
work.

We measure the brightness of stars as well as their location from the
constructed images. We use a series of DAOPHOT tasks to find stars and
perform PSF fitting photometry with the same PSF that we used to
construct the images. Since Poisson noise is included when
constructing the images, this does not make the subtraction
perfect. The process produces an image where all the stars have been
subtracted and only the background light remains. We call these images
`subtracted', while we refer to the original unsubtracted image as
`full'. We measure a surface brightness profile in concentric annuli
from the center of the cluster by calculating the biweight of the
counts in an annulus and dividing by the number of pixels in that
annulus. We use two sets of annuli for these measurements. The first
set goes from 1 to 25 pixels in steps of $4-6$ pixels, and the second
set goes from 20 to 100 in steps of 20 pixels. The assigned radius of
a given annulus is the average between the inner and outer radii. In
paper~I we note that, for the input profiles with steep cusps, the
subtracted images produce a flatter central profile than the
input. This is because crowding produces and over subtraction after
the PSF fitting process. As done in paper~I, we decide to produce
alternative images with only the 10\% brightest stars PSF subtracted
and an image where we mask the 2\% brightest stars with a radius of 3
pixels to try to avoid the over subtraction problem.

Once we have the catalog of found stars we can compare it to the
original input list and estimate how many of the input stars are found
for different magnitude a radial bins. For the cases with 200,000
input stars We find that bright stars are found with an efficiency
higher than 100\%, meaning that more stars are found to be in the
brightest magnitude bin than the number of stars that were input for
that same magnitude bin. This effect is more pronounced in the central
part of the cluster, where crowding problems are worse. The effect is
easy to explain since one or more faint stars are likely to fall
within the PSF disk of the bright stars and their light is measured as
if it was part of the bright stars. The difference between input and
measured magnitudes is typically $\sim$0.1 mag, which is enough to
push some stars from a fainter magnitude bin to a brighter one. For
intermediate-magnitude stars, the same effect happens, for the regions
near the core; the efficiency for finding these stars falls to
$50-70$\% depending on the shape of the profile. The efficiency for
finding the fainter stars is lower in any radial bin; it is a few
percent in the center and up to 50\% for the regions at large
radii. As expected, these numbers become more extreme for the cases
with steeper central slopes, since crowding is worse then. For the
case with 50,000 input stars, the trends are similar, but the numbers
are less extreme. Stars in the brightest magnitude bin are found with
an efficiency close to 100\% for the cases with flatter central
slopes. The efficiencies for the cases with steeper central slopes are
very similar to those with 200,000 input stars. Finally, for the cases
with 10,000 input stars. The efficiencies for finding the input stars
are all close to 100\% except for the faintest stars in the central
region of the cluster, which are around $70-80\%$ depending on how
steep the central slope is. The conclusion from this analysis is that
when correction factors are calculated for star count measurements,
the factors are dependent on the shape of the density profile and the
number of existing stars. If one assumes the wrong shape or the wrong
number of stars in the cluster, the correction factors will be
incorrect.

Stars are counted in and divided by the area of each annulus. The
above discussion about the efficiency for finding stars implies that
the stars below a certain brightness are never found with 100\%
efficiency, therefore we exclude them from the star lists. We compare
the obtained star count profiles with those obtained by measuring
integrated light from the four different images (full, subtracted,
10\% subtracted and 2\% masked). Results from these measurements are
shown in Figure~\ref{ch2f2} for the 200,000 input stars case and
Figure~\ref{ch2f3} for the 50,000 input case. In both figures we are
showing models 1, 4, and 5, which have central slopes of 0, $-0.4$ and
$-0.7$ respectively. We find that, depending on the shape of the input
model, the profile measured in the subtracted, partially subtracted,
or masked images follow the input profile best. For the least
concentrated cases, the measurements from the subtracted and 10\%
subtracted image seem to follow the profile best, but for the more
concentrated cases, the subtracted and 10\% subtracted cases tend to
look flatter in the center than the input profile. For these cases,
the profile from the masked image seems to be a better choice. The
star counts profiles are always much noisier than the light profiles
in the central regions and they show a consistent bias in the central
regions for the cases with steep central slopes.

\begin{figure*}[t]
\centerline{\psfig{file=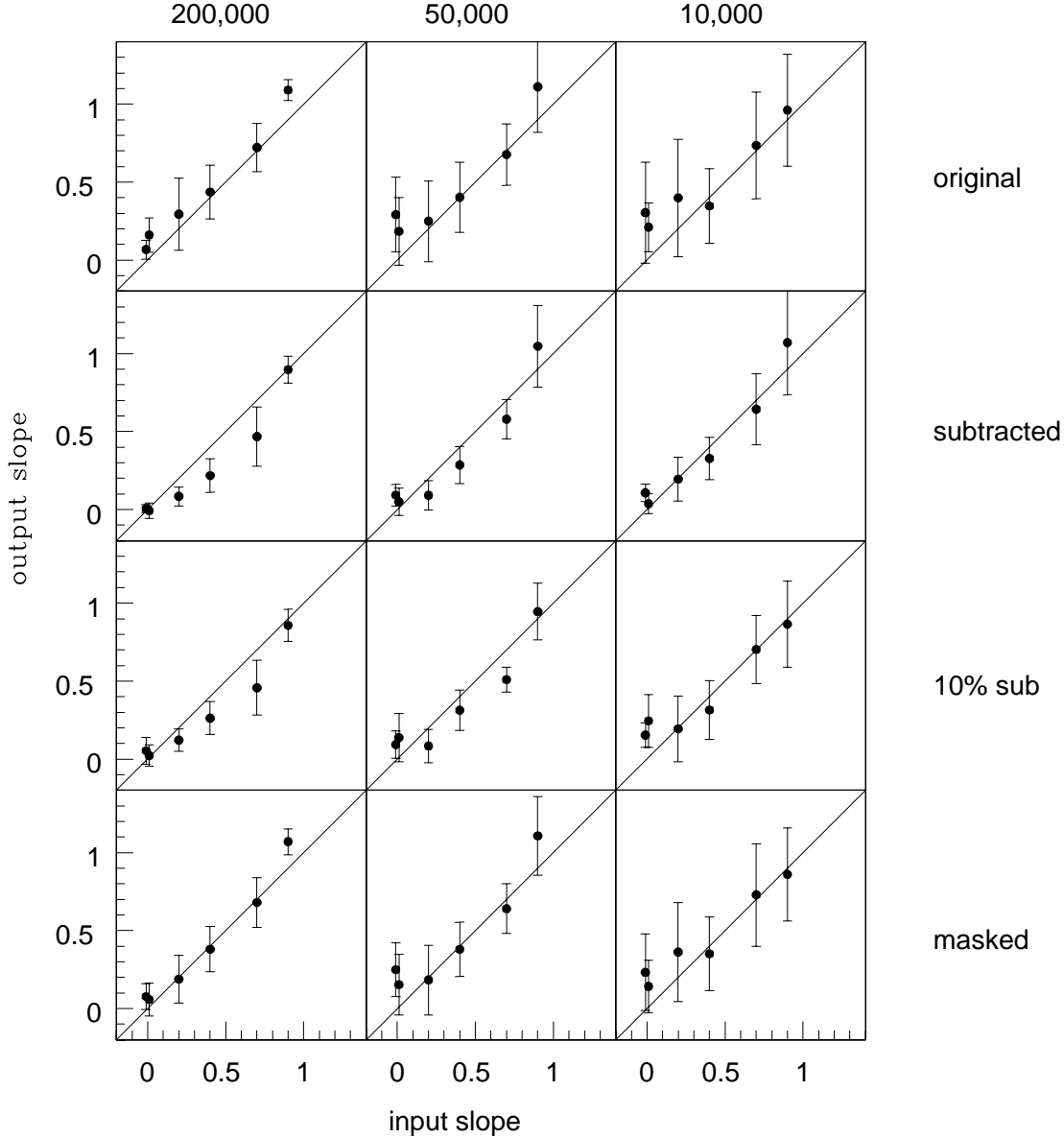,width=16.5cm,angle=0}}
\figcaption[Input vs. measured central slopes for the simulated
  datasets]{Input versus measured surface brightness slope for the
  different groups of simulations. A small horizontal offset is
  introduced for clarity in the case of the two models with central
  zero slopes. We show the average measured slope of the individual
  profiles for each case. Error bars represent one standard deviation
  for all the measurements.}
\label{ch2f4}
\end{figure*}

\begin{deluxetable*}{lllllllllllll}
\tablewidth{0pt}
\tabletypesize{\scriptsize}
\tablecaption{\label{tbl2b}LMC, SMC and Fornax Sample.}
\tablehead{
\colhead{name} &
\colhead{parent} & 
\colhead{primary} &
\colhead{filter} &
\colhead{exp. time} &
\colhead{secondary} &
\colhead{filter} &
\colhead{exp. time} &
\colhead{$\alpha$ center} &
\colhead{$\delta$ center} & \\
\colhead{} & 
\colhead{galaxy} &
\colhead{dataset} &
\colhead{} &
\colhead{(sec)} &
\colhead{dataset} &
\colhead{} &
\colhead{(sec)} &
\colhead{} &
\colhead{} &
}
\startdata
NGC~1466 & L & U2XJ0105B & F555W & 3520 & U2XJ0108B & F814W & 4520 & 03:44:32.75 & -71:40:16.53  \\
NGC~1651 & L & U2S75801B & F555W & 1000 & U2S75803B & F814W & 1000 & 04:37:31.95 & -70:35:06.99  \\
NGC~1711 & L & U2Y80501B & F555W & 1520 & \nodata & \nodata & \nodata & 04:50:39.95 & -69:59:06.60  \\
NGC~1754 & L & U2XQ0103B & F555W & 1540 & U2XQ0109B & F814W & 1860 & 04:54:18.35 & -70:26:31.60  \\
NGC~1786 & L & U2XJ0205B & F555W & 3520 & U2XJ0208B & F814W & 4520 & 04:59:07.58 & -67:44:44.96  \\
NGC~1805 & L & U4AX0204B & F555W & 435  & U4AX020AB & F814W & 960  & 05:02:21.48 & -66:06:41.60  \\
NGC~1818 & L & U4AX3603B & F555W & 2500 & U4AX3703B & F814W & 2500 & 05:04:10.58 & -66:26:26.63  \\
NGC~1835 & L & U2XQ0203B & F555W & 1540 & U2XQ0209B & F814W & 1860 & 05:05:06.97 & -69:24:13.28  \\
NGC~1866 & L & U5DP020TB & F55W5 & 2512 & U5DP020PB & F814W & 2620 & 05:13:29.00 & -65:27:15.37  \\
NGC~1868 & L & U4AX5803B & F555W & 2500 & U4AX5903B & F814W & 2500 & 05:14:34.53 & -63:57:14.68  \\
NGC~1898 & L & U2XQ0303B & F555W & 1540 & U2XQ0309B & F814W & 1860 & 05:16:41.92 & -69:39:23.96  \\
NGC~1916 & L & U2XQ0403B & F555W & 1549 & U2XQ0409B & F814W & 1860 & 05:18:37.79 & -69:24:27.05  \\
NGC~1984 & L & U5AY0901B & F555W & 1410 & U5AY0904B & F814W & 1410 & 05:27:39.97 & -69:08:02.14  \\
NGC~2004 & L & U2Y80201B & F555W & 1520 & U2Y80204B & F814W & 1510 & 05:30:40.24 & -67:17:15.64  \\
NGC~2005 & L & U2XQ0503B & F555W & 1540 & U2XQ0509B & F814W & 1860 & 05:30:10.32 & -69:45:08.82  \\
NGC~2019 & L & U2XQ0603B & F555W & 1540 & U2XQ0609B & F814W & 1860 & 05:31:56.47 & -70:09:32.48  \\
NGC~2031 & L & U2Y80301B & F555W & 1520 & U2Y80304B & F814W & 1510 & 05:33:40.70 & -70:59:07.44  \\
R136 & L & U2HK030JB & F555W & 1211 & U2HK0317B & F814W & 1205 & 05:38:42.52 & -69:06:02.98  \\
NGC~2100 & L & U5AY0701B & F555W & 1410 & U5AY0704B & F814W & 1410 & 05:42:07.66 & -69:12:43.47  \\
NGC~2214 & L & U5AY1101B & F555W & 1410 & U5AY1104B & F814W & 1410 & 06:12:56.92 & -68:15:37.92  \\
NGC~2257 & L & U2XJ0505B & F555W & 3520 & U2XJ0508B & F814W & 4520 & 06:30:00.89 & -64:19:23.26  \\
KRON 3 & S & U26M0G02T & F555W & 300 & \nodata& \nodata & \nodata & 00:24:46.03 & -72:47:35.09  \\
NGC~121 & S & U3770501B & F555W & 1840 & U377050BB & F814W & 2080 & 00:26:48.62 & -71:32:09.10  \\
NGC~330 & S & U5AY1001B & F555W & 1410 & U5AY1004B & F814W & 1410 & 00:56:18.41 & -72:27:49.65  \\
NGC~411 & S & U26M0302T & F555W & 300 & \nodata& \nodata & \nodata & 01:07:56.35 & -71:46:01.59  \\
NGC~416 & S & U26M0502T & F555W & 200 & \nodata& \nodata & \nodata & 01:07:59.30 & -72:21:17.43  \\
FORNAX 2 & F & U30M020EB & F555W & 5640 & U30M020IB & F814W & 7720 & 02:38:44.26 & -34:48:27.11  \\
FORNAX 3 & F & U30M030EB & F555W & 5518 & U30M030IB & F814W & 7720 & 02:39:48.22 & -34:15:26.87  \\
FORNAX 4 & F & U2LB0203B & F555W & 2400 & U2LB0205B & F814W & 2400 & 02:40:09.01 & -34:32:19.81  \\
FORNAX 5 & F & U30M040EB & F555W & 5640 & U30M040IB & F814W & 7720 & 02:42:21.14 & -34:06:04.32  \\
\enddata
\end{deluxetable*}

We test how well we recover the input central slope for the different
shapes of input profiles and for the different measurement
methods. Since we measure the central slope by taking a first
derivative of the profile we need a smooth version of it. For this, we
apply the one dimensional version of the spline smoother mentioned in
section~\ref{scen}. This allows to recover information from the
profile without fitting any parametric model to the data. We exclude
the star count profiles from these measurements because the central
parts of the profiles are too noisy for the spline smoother to get a
reasonable fit. The first derivative of the smooth profile has a
section toward the center where it is constant; we take this constant
value as the measured central slope. After measuring the central slope
for the different realizations, we calculate the average and the
standard deviation for each case. We show the input versus measured
central slopes in Figure~\ref{ch2f4}.

We observe that the uncertainty on the slope measurements increases as
the number of input stars decreases. For the 10,000 input stars case,
the profiles from the subtracted and 10\% subtracted images yield
smaller uncertainties. These two cases tend to underestimate the
central slopes for the concentrated and rich (200,000 and 50,000 input
stars) cases, while the slopes recovered from the masked images seem
to follow the input better. For all the rich cases, the measurements
for the model with the steepest central slope overestimate the slope,
we think this can be due to the fact that so many stars are being
input at the center that not enough stars are being input for the
outer parts, which would explain the fact that we find fewer stars for
this case.

We test our ability to measure the input break radius by measuring the
minimum of the second derivative, which is the radius at which the
curvature is maximum. Our results show that we can measure the break
radii for the simulated clusters to within 10\% accuracy. The majority
(all except two) of the observed clusters have a reported core radius
larger than the one for our simulations, so we are confident that we
can measure such break radii.

\subsection{Uncertainties}\label{sunc}

We refer the reader to the detailed discussion in section~2.4 of
paper~I about the sources for uncertainty when measuring surface
density profiles from integrated light versus measuring it from star
counts. In order to properly estimate our uncertainties, we compare
the photometric scatter between different realizations having
identical input parameters with the biweight scatter estimate. In
paper~I we find that the biweight scatter has to be scaled in order to
match the photometric scatter measured from the different
realizations. For these new simulations we find that the scaling
factors change due to the differences in our input shapes and of
simulating clusters at larger distances (the number of stars on a
given annulus and differences in PSF). As done in paper~I, we compare
these scaling factors with those obtained for real data from an
alternative method discussed in section \ref{dunc}.

We also estimate the error in our central slope measurements by
comparing the scatter of measured slopes with the known input slope
for every simulated cluster. The results are shown in
Fig~\ref{ch2f4}. We confirm what we learned from analyzing
Figs~\ref{ch2f2} and \ref{ch2f3}. The slope uncertainties are smaller
for the subtracted and partially subtracted cases, but they are biased
low for the cases with steep cusps and large number of input
stars. Also, the slope measurements are more uncertain for the
clusters with 10,000 input stars. The figure suggests to take the
measurements from the masked image for the cases with steep central
profiles and the measurements from the subtracted or 10\% subtracted
for the others. In the case of 10,000 input stars, the subtracted case
always seems to be better and is not biased.

\section{Data and Analysis}\label{data}

\subsection{Sample}\label{sam}

As mentioned in paper~I, there are minimum requirements for an image
to be suitable for measurements with our technique. The image needs to
have a minimum number of counts, which can be obtained by having a
large number of stars present due to richness, high concentration, or
by having long exposure times. We establish that detecting stars six
magnitudes fainter than the horizontal branch with a signal to noise
of 20 is a minimum requirement for low-concentration clusters. This
criterion can be relaxed for highly concentrated clusters and for
those with a large number of stars ($M_V<-7.5$). Taking into account
these requirements, we gather 30 clusters from the $\it{HST}$
archive. The sample contains 21 clusters in the LMC, 5 in the SMC and
4 in the Fornax dwarf galaxy. When images are available in two filters
(F555W and F814W), we align and combine the images in order to improve
signal to noise. We believe we are justified in doing this because the
color gradients for the radial range that we are measuring are smaller
than the photometric uncertainties. If no alternative image is
available, we use the single F555W dataset. In general, we analyze
only the chip in which the center of the cluster lies, the only
exception is the cluster Kron~3, for which we use all four chips. The
size of one WFPC2 chip is large enough to contain a few core radii for
every cluster in the sample. The scale of the CCD is 0.1\arcsec/pixel
for the WF chips and 0.046\arcsec/pixel for the PC chip.

We use the WFPC2 associations from the Canadian Astronomy Data Center
website\footnote{http://cadcwww.dao.nrc.ca/}. These images are spatial
associations of WFPC2 images of a given target. The raw data frames
are processed through a standard reduction pipeline, grouped in
associations and combined. The available data are a multi-group image
with frames for the three WF and the PC chips. It is straightforward
to align and combine two of these images from different filters if
they belong to the same program, which is the case for every object
with two images available in our sample.

\begin{figure*}[t]
\centerline{\psfig{file=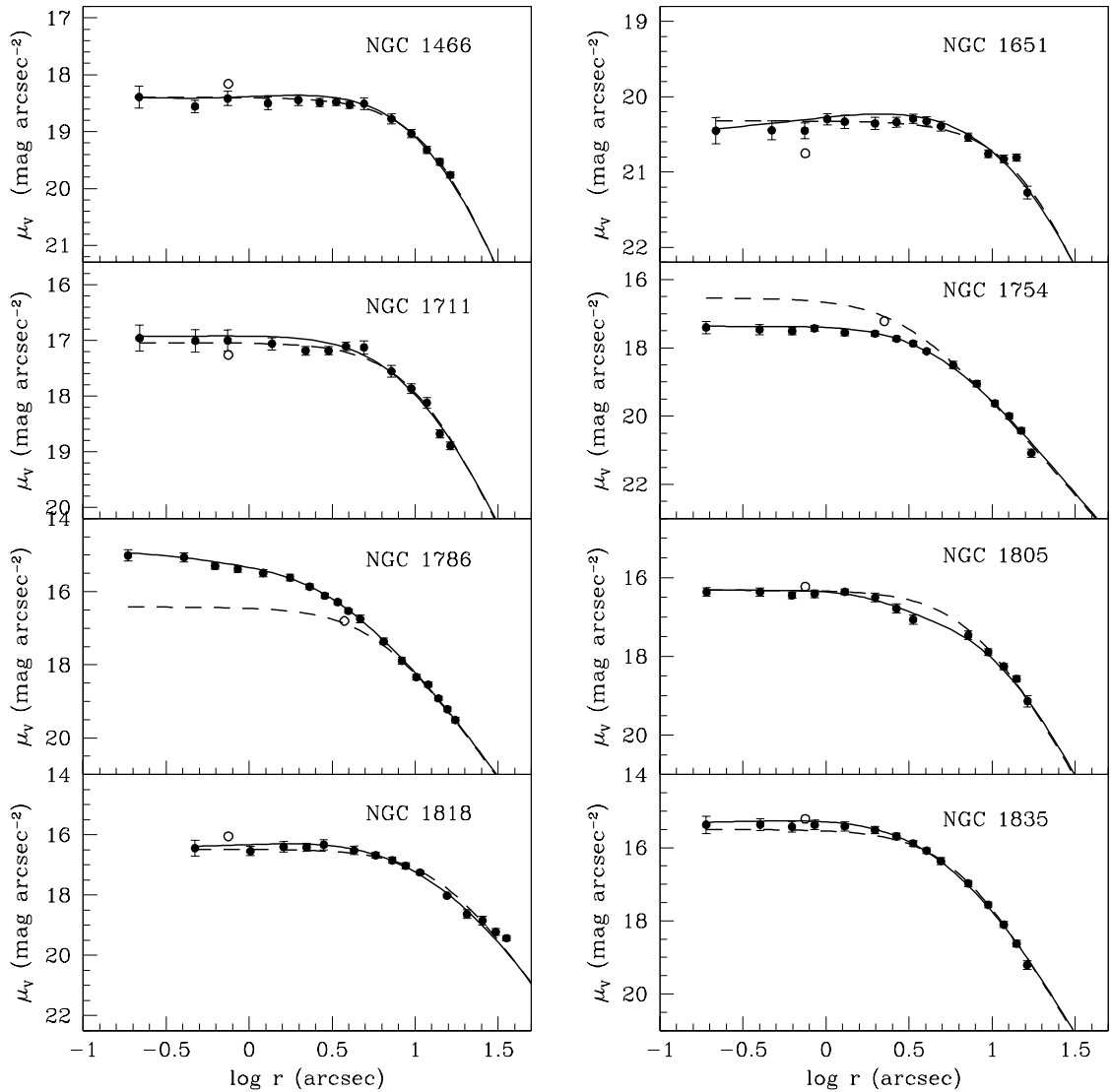,width=16.5cm,angle=0}}
\figcaption[SB profiles for the LMC globular clusters]{Surface brightness
  profiles for the LMC clusters. For each cluster we show our
  photometric measurements (solid points), our smooth profile (solid
  line), and the EFF fit by MVM05 (dotted line). The smooth profile
  comes from a fit to our photometric points inside $\sim10$\arcsec\
  and the EFF fit outside that region. For every panel the SB units
  are V mag/arcsec$^2$.}
\label{ch2f5a}
\end{figure*}

\setcounter{figure}{4}

\begin{figure*}[t]
\centerline{\psfig{file=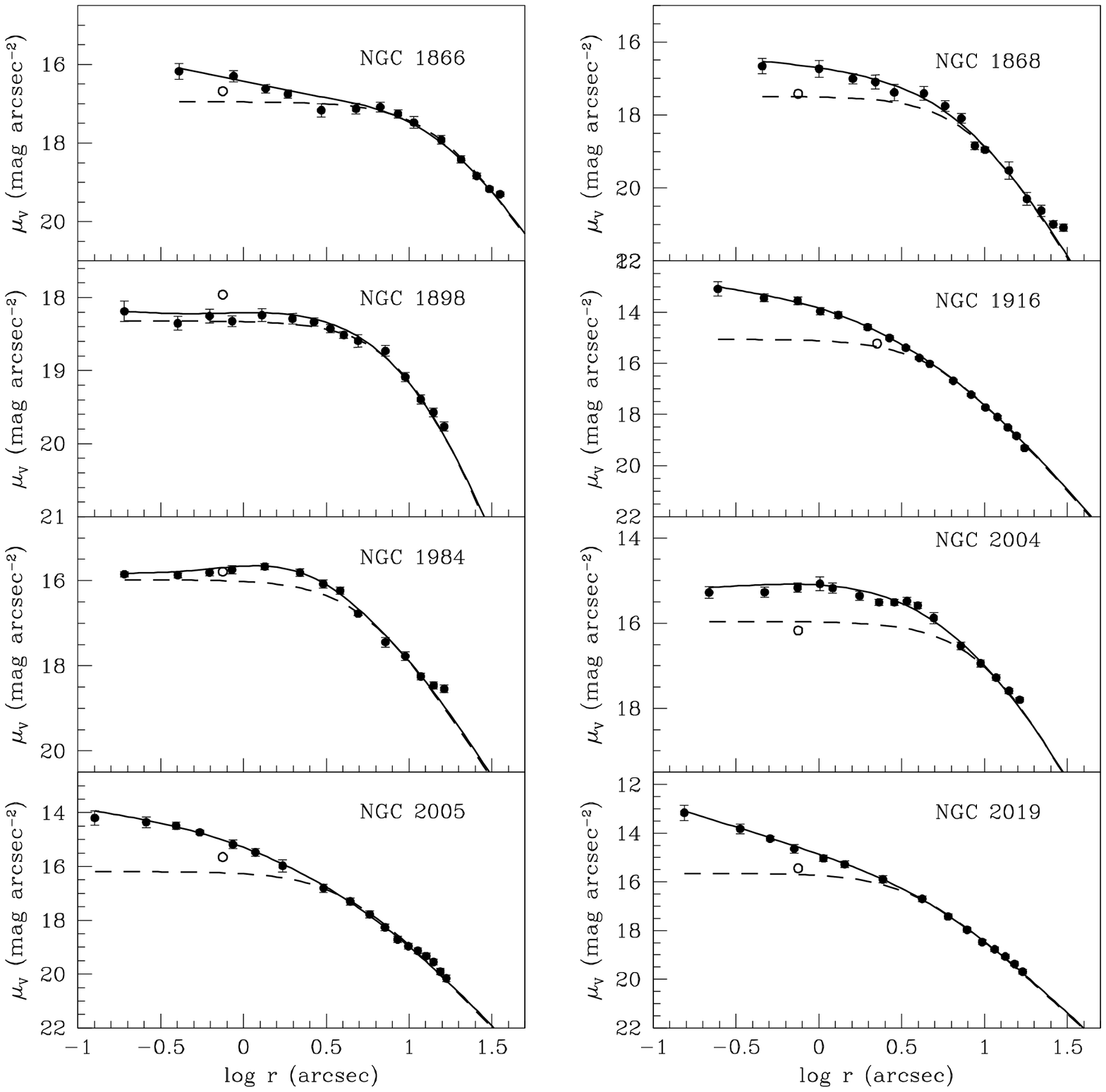,width=16.5cm,angle=0}}
\figcaption[SB for LMC continued]{continued}
\label{ch2f5b}
\end{figure*}

\setcounter{figure}{4}

\begin{figure*}[t]
\centerline{\psfig{file=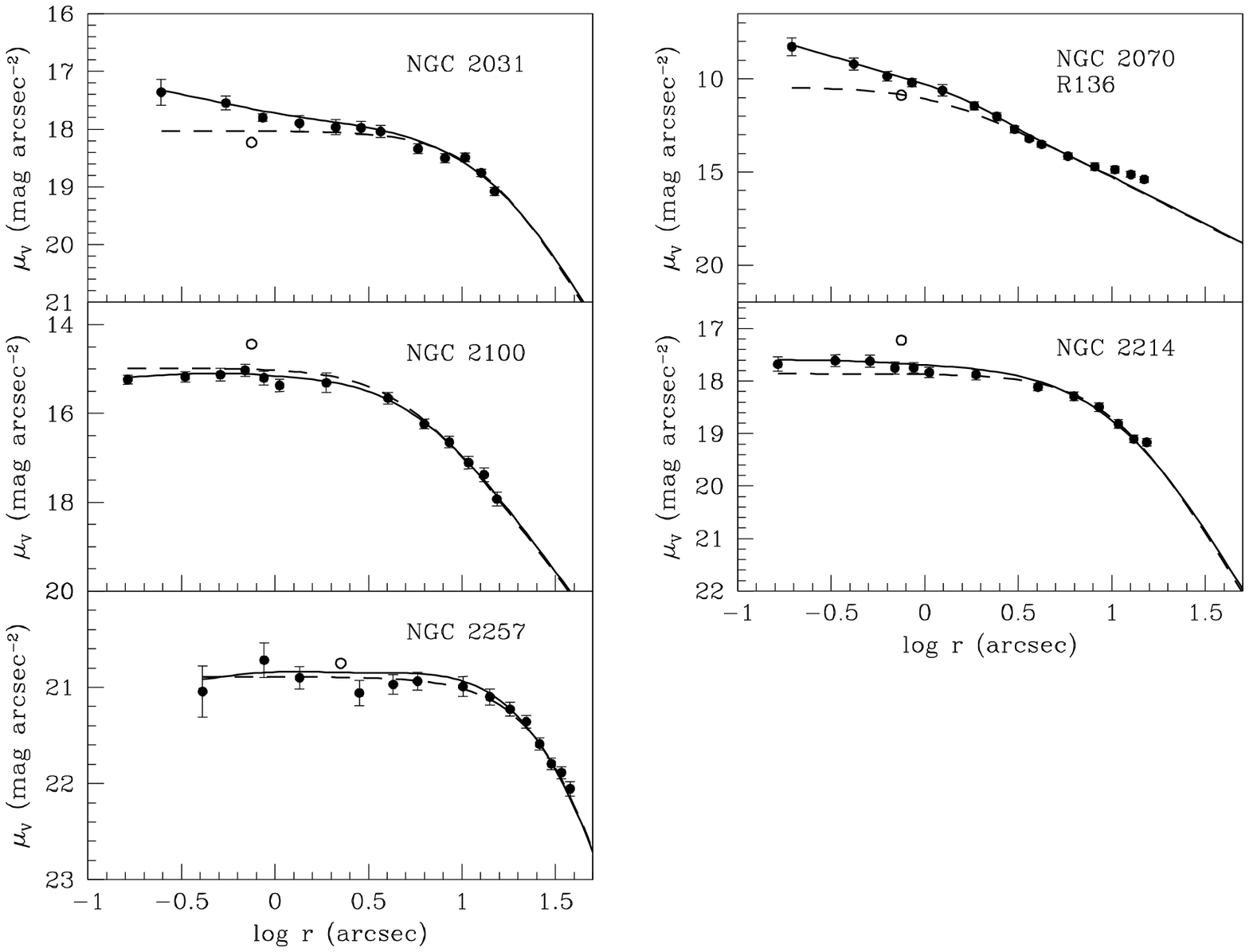,width=16.5cm,angle=0}}
\figcaption[SB for LMC continued]{continued}
\label{ch2f5c}
\end{figure*}

\begin{figure*}[t]
\centerline{\psfig{file=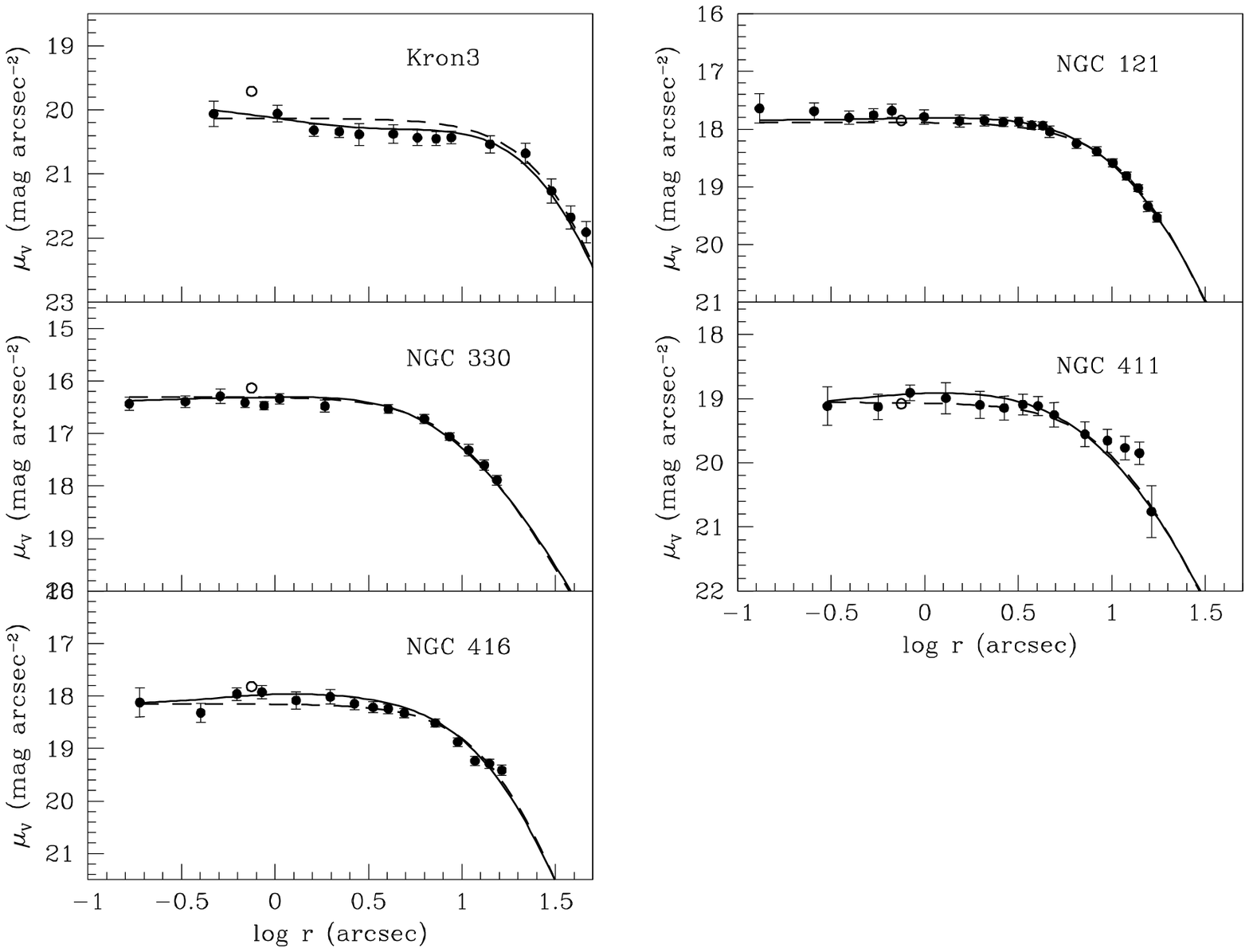,width=16.5cm,angle=0}}
\figcaption[SB profiles for SMC globular clusters]{The same as in Fig
\ref{ch2f5a} for the SMC clusters.}
\label{ch2f6}
\end{figure*}

\begin{figure*}[t]
\centerline{\psfig{file=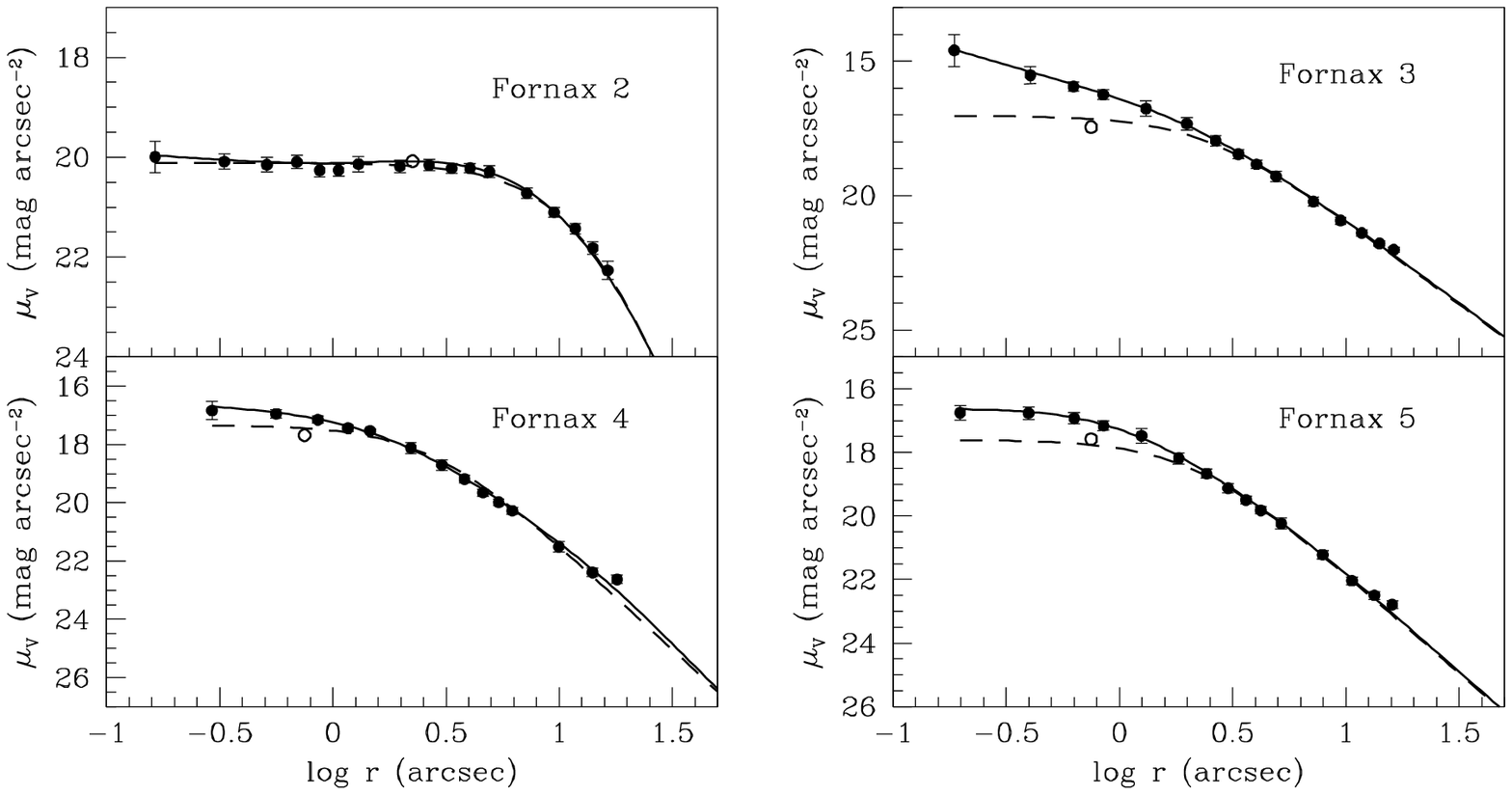,width=16.5cm,angle=0}}
\figcaption[SB profiles for the Fornax dwarf galaxy globular
  clusters]{The same as in Fig \ref{ch2f5a} for the Fornax dwarf galaxy
  clusters.}
\label{ch2f7}
\end{figure*}

\subsection{Image Processing}\label{dim}

We process the data in the same way we do for the simulated images. We
choose the frame where the cluster center is located, this is usually
the PC frame, but for a few cases, it is one of the WF frames. We trim
the image in order to eliminate the noisy edges and then proceed to
perform basic photometry with various DAOPHOT tasks. First, we use the
``FIND'' task to make a preliminary list of detected stars, then we
perform aperture photometry with the task ``PHOT'' in order to choose
candidates for PSF construction. We find that PSF stars have to be
chosen by hand because a single bad PSF star can have an important
effect in the final PSF construction. Once we have a list of PSF
stars, we perform an iterative procedure in which we subtract the
neighbor stars to the PSF stars and then recalculate the PSF. In this
way, the PSF construction is less affected by crowding. Using the
final constructed PSF we subtract all the stars from the image,
leaving behind an image containing only background light. We also
create an image with the brightest 10\% stars subtracted and another
one with the brightest $\sim$2\% stars masked.

Some of the data frames contain a small number (2 to 6) of very bright
stars that appear saturated in the images. These bright stars are
found at various locations on the image, but the closer they are to
the center, the larger the effect they can have on our measurements by
creating bumps in the profile. We decide to exclude these stars from
our measurements by masking them with a larger masking radius than the
one used for the 2\% brightest stars. The clusters for which this
extra step was taken are NGC~1818, NGC~1984, NGC~2100, NGC~2214, and
NGC~330.

\vspace{10pt}

\subsection{Center}\label{dcen}

As discussed in paper~I, it is crucial to measure the center of the
cluster accurately in order to measure a reliable surface brightness
profile. We use the method outlined in section \ref{scen} to estimate
the center coordinates for each cluster. Many clusters in the LMC are
known to have an elliptical shape. Our method is not affected by this
as long as the cluster is symmetric in two dimensions. The first guess
center is always chosen by eye, and the radius for counting stars is
chosen so that the circle lies entirely inside the image, therefore,
it is larger if the center is closer to the middle of the chip. For
one cluster (NGC~1868), the center of the cluster is very near the
edge of one of the chips. In this case, we calculate the center from a
different image with lower exposure time, but with the center located
in the middle of the chip. We report our measured centers in
table~\ref{tbl2b}. We warn the reader that special care should be
taken when using these coordinates. These centers are valid only using
the world coordinate system (WCS) information contained in the header
of each image. The WCS information for two different images can make
the coordinates for same location vary by a few arcsec. The center is
always measured on the primary dataset (the F555W image) when two
images were combined.

\subsection{Surface Brightness Profiles}\label{dsb}

Once we have measured a center, we calculate the surface brightness
profile from the four different images of each cluster. We calculate
surface brightness by estimating the biweight (as explained in
section~\ref{ssb}) of the number of counts per unit area in a series
of concentric annuli. The choice of the size of annuli in which we
measure the profile is given by a trade off between spatial resolution
and noise. For images with very high signal to noise, we can use
smaller steps, while for more sparse cases, smoother profiles are
obtained by increasing the size of the bins at the cost of decreasing
the spatial resolution. We use three different sets of annuli, the
first one goes from 1-25 pixels with steps of 4-6; the second goes
from 20-100 pixels with steps of 12-15; and the third one from 100-380
pixels with steps of 40-60.

In section~\ref{ssb} we observe that for the simulated images the star
counts profile tends to underestimate the profile at the center and it
is noisier than the integrated light profile. For this reason, we
decide not to calculate the star count profiles for these
datasets. Also, we observe that the profiles coming from the
unsubtracted image are always noisier than those obtained from the
other images, so we never use the `full' profile as our final
result. For every set of simulations, the subtracted and 10\%
subtracted images always yield smoother profiles, unfortunately they
show to be biased toward the center for the profiles with steep
central slopes, so we can only use them when all four profiles are
consistent with a central flat profile. If there are systematic
differences between the original and masked profiles and the two
subtracted ones in the sense that the first two are steeper than the
latter two, then we use the profile from the masked image, since this
is the one that traces the central cusp best in our simulations.

In paper~I we find the photometric zero point by integrating our
measured light profiles and comparing them to previously obtained
profiles from ground based data. We cannot do the same thing here
because our profiles have a smaller radial extent. For the cases in
which our central profile differs significantly from previous
measurements, the radial extent in which the two profiles agree is not
large enough for us to make a meaningful comparison of enclosed
light. We also observe that the differences in shape between our
measurements and those obtained by MAC03 are always inside the
turnover radius. We therefore use the data points outside the core
radius to normalize our profiles to the EFF fits by MVM05. We choose
to normalize to these profiles because MVM05 use the MAC03 photometric
points, but they re-normalized them using ground based data and they
correct for reddening. This brings all our measurements to a common
scale on V$_{mag}$/arcsec$^2$.

We want to make measurements of central and outer slopes, but our
images are radially limited, so we construct radially extended
profiles by using our measured profile inside $\sim10$\arcsec\ and the
MVM05 EFF fits outside that radius. We measure slopes by taking a
first derivative of the profiles, which requires a smooth version of
the profiles since noise is greatly amplified when taking
derivatives. The smoothing is done by using the one dimensional spline
mentioned in section~\ref{ssb} \citep{wah90}. For most profiles, there
is a region in the center for which the first derivative is
constant. We take this value as the central slope. For the clusters
that show a steep central cusps, the slope sometimes changes through
the entire radial range. In this case we take the central most value
of the derivative as the inner slope. Since we use EFF fits at large
radius, we expect the first derivative to reach a constant value
outside. The measured value is expected to coincide with the slope of
the power-law for the EFF fits, which it indeed does. We take the
value of the first derivative at the half-light radius as the outer
slope. It is worth noting that the measured outer slope will be very
different to that measured for Galactic clusters, since the profiles
for those clusters are calculated from a King fit, which does not have
a constant outer slope.

We deproject the profiles after smoothing in order to obtain a
luminosity density (LD) profile for each cluster. For the clusters
with flat central profiles we often cannot obtain a proper
deprojection due to noise, because the noise sometimes makes the
central points be slightly fainter than the rest, which produces a
positive slope in the smooth profile and that cannot be deprojected
numerically. For the cases in which we do obtain a proper
deprojection, we measure the central slope of the LD distribution in
the same way as we measure the central SB slope, by taking a first
derivative.

The traditional measurement of core radius as the radius for which the
central luminosity value falls by half loses meaning when the central
slope is not zero. If the central profile shows a cusp, then the core
radius will be resolution-dependent. For this reason, we decide to
measure what we call a break radius instead. The break radius is
defined as the radius of maximum curvature, the one in which the
second derivative reaches a minimum. This is a more systematic measure
for a set of non-parametric profiles with different central
slopes. Even after applying the smoothing procedure, there is still a
certain amount of noise present in the second derivative, for this
reason, we fit a high order polynomial and take the minimum of the fit
instead of the minimum of the second derivative as our break
radius. In paper~I we find that the core and break radius coincide
for the clusters having a flat core, but the do not coincide for the
cases presenting a cusp.

The difference in shape from our measurements and the parametric fits
will affect the measurement of the half light radius. Since we are
using EFF fits for the outer part, and these fits are formally
infinite, we have to truncate them in order to measure the total
enclosed light. We use the tidal radius measured by MVM05 as a
truncation radius and measure the half light radius for our smooth
profiles. Having an estimate of the total luminosity and using the M/L
values calculated by MVM05 we can estimate the total mass of each
cluster and thus estimate the median relaxation time as described in
\citet{bin87}

$$ t_{rh}=\frac{2.06\times10^6}{\ln(0.4M_t/\langle m\rangle)}\langle m\rangle^{-1}M_t^{1/2}r_h^{3/2} .$$ 

\vspace{5pt}

\noindent We assume a mean mass of $0.5M_\odot$ as in MVM05. Results
from these calculations are presented in table~\ref{tbl3b}.

\subsection{Data Uncertainties}\label{dunc}

We describe how we estimate uncertainties for the simulations in
section~\ref{sunc}. The method is based on different realizations for
which shot noise from stars can be estimated directly. We use an
alternative method to calculate the uncertainties for real data and we
calibrate this method against that used for the simulations, as we did
in paper~I. We assume a smooth underlying stellar radial profile, so
the uncertainties of the photometric points should reflect deviations
from a smooth curve in a statistically meaningful way (i.e., have a
Gaussian distribution around the mean value). We calculate the root
mean square (RMS) difference between the smooth profile and the data
points for the central region. The biweight yields an estimate for the
central location (SB value) and scale (scatter); this scale value is
divided by the square root of the number of sampled pixels and used as
the initial uncertainty for individual photometric points. We then
calculate the ratio of the biweight to the RMS, which should represent
our lack of inclusion of shot noise from the stars. This ratio depends
on the extent of the radial bins (i.e, the number of pixels used),
therefore we use different scalings for the different binnings. We
estimate the scales for the simulations using the different
realizations, in order to make sure that the scalings coincide. The
average scaling for the inner points is about~3 and about~10 for the
outer points. These numbers are consistent with what we found in the
simulations. Thus, we are effectively including shot noise from
stars. The largest scalings occur for sparse clusters, as expected.

In the same way as in paper~I, we calculate the uncertainties on
slope measurements from a bootstrap technique. The bootstrap approach
follows that in \citet{geb96}. From the initial smooth profile, a new
profile is created by generating random values from a Gaussian
distribution with the mean given by the initial profile and the
standard deviation from the photometric uncertainties. A hundred
profiles are generated in this way and the $16-84\%$ quartiles are
measured for the errors. These estimated errors are compared with the
scatter measured for the simulated cases in fig~\ref{ch2f4} and the
two independent error measurements agree quite well, which gives us
the confidence that the uncertainties calculated with the bootstrap
method are reliable. In paper~I we perform one more check on our
slope uncertainties by measuring the effect of increasing the
uncertainties on photometric points by a factor of two. From the
bootstrap method, we find that the slope uncertainties increased by a
modest factor, less than two, for most clusters. Thus, the slope
uncertainties are not too sensitive to individual photometric errors.

\vspace{20pt}

\section{Results and Discussion}\label{res}

\subsection{Surface Brightness}\label{rsb}

The measured surface brightness profiles for the entire sample are
shown in figs~\ref{ch2f5a} to~\ref{ch2f7}. For each cluster we show
our normalized photometric points with error bars, and a smooth
profile made from the combination of our photometric points inside
$\sim$10\arcsec\ and MVM05 EFF fits outside that radius. For
comparison we show the MVM05 EFF fit and the central photometric point
used for that fit. We would like to stress that our measured
photometric points at radii larger than $\sim$10\arcsec\ do not
participate in the construction of the smooth fit, instead the EFF
fits are used in that region. For about half the sample (17 objects),
the agreement between our measurements and the EFF fits of MVM05 is
excellent, even for those cases in which the central photometric point
by MAC03 is barely inside the turnover radius (such as Fornax~2) or it
does not lie on top of the EFF fit (such as NGC~1651, NGC~1898, or
2100). There is only one case (NGC~1754) for which our photometric
points are fainter than the EFF fit. For the remaining 12 objects, our
photometric points are brighter than the EFF fit by more than 0.5
mag/arcsec$^2$, with three objects (NGC~2019, R136, and Fornax~3)
having differences larger than 2 mag/arcsec$^2$.

MAC03 identify a few clusters that they think agree with the expected
post core-collapse (PCC) morphology by showing a power-law cusp in
their central profile. NGC~2005 and NGC~2019 are identified as clear
cases of PCC morphology with central power-law slopes of
$-0.75$. NGC~1835 and NGC~1898 are marked as good candidates for PCC
morphology, but they measure lower power-law slopes of $-0.45$ and
$-0.30$ for them. Three more clusters, NGC~1754, NGC~1786, and
NGC~1916 have incomplete profiles and are classified as intriguing due
to their small cores, but are not placed as firm PCC
candidates. Fornax~5 is also considered a good candidate for a PCC
cluster based on it's small core and central profile shape. Our
results for these seven clusters confirm the presence of a steep cusp
for NGC~2005, NGC~2019, and NGC~1916; and a shallow cusp for NGC~1786.
The rest of the cases all show flat central cores. Our reported values
of the central slopes are different from the power-law slopes of
MAC03, this makes sense since they are fitting a power-law to
photometric points on a larger radial range than that in which we are
measuring central slopes. 

Independently of PCC morphology, we identify a few more clusters as
having clear central cusps (with central slopes steeper than $-0.20$)
such as NGC~1866, NGC~2031, Fornax~3 and Fornax~4, and some showing
shallow cusps (with central slopes flatter than $-0.20$) such as
NGC~1868, NGC~2214, and Fornax~2. When the luminosity density central
slopes are taken into account, a similar classification arises,
NGC~1866, NGC~1916, NGC~2005, NGC~2019, and Fornax~3 have steep cusps
with LD logarithmic slopes steeper than $-1.00$, while NGC~1754,
NGC~1868, NGC~2031, and Fornax~4 show shallow cusps with slopes
between $-0.2$ and $-1.0$. The cluster R136 is discussed in a separate
section (\ref{r136}). We should note that some of these clusters have
half-light relaxation times longer than their measured age, they
cannot be expected to have undergone core-collapse and therefore,
another mechanism has to be invoked to explain the central non-zero
slopes. \citet{bas06,goo06} suggest that young star clusters can be
out of Virial equilibrium due to rapid gas losses, and therefore, the
shape of their surface brightness profiles can change on relatively
short timescales.

\vspace{20pt}

\subsection{R136}\label{r136}

R136 is known to be an extremely young object at the center of the 30
Doradus nebula in the LMC. It is considered to be a young version of a
globular cluster due to its large content of O type stars. Main
sequence stars with masses as high as $120 M_\odot$ have been detected
in it \citep{mas98}. The estimated age for the most massive stars is
$<1-2$ Myr and the mass function agrees very well with a Salpeter
initial mass function (IMF). This makes R136 a unique and very
peculiar object because it allows us to study star clusters in the way
they looked just after formation. The surface brightness profile that
we measure has a logarithmic central slope that is steeper than
anything measured before for a globular cluster and steeper than
anything predicted by dynamical models like core-collapse. This makes
us suspect that we are not resolving a core or a turnover radius for
this object and that our central slope measurement corresponds to the
slope just outside the turnover radius for the other objects. We
decide to include R136 in every systematic measure we made for other
clusters, but we caution the reader that its location in different
distributions, particularly those dealing with central SB slope,
should be taken with a grain of salt for this reason. The central
surface brightness value for this object implies a central density of
$8\times10^6M_\odot/pc^3$.

\begin{figure}[t]
\centerline{\psfig{file=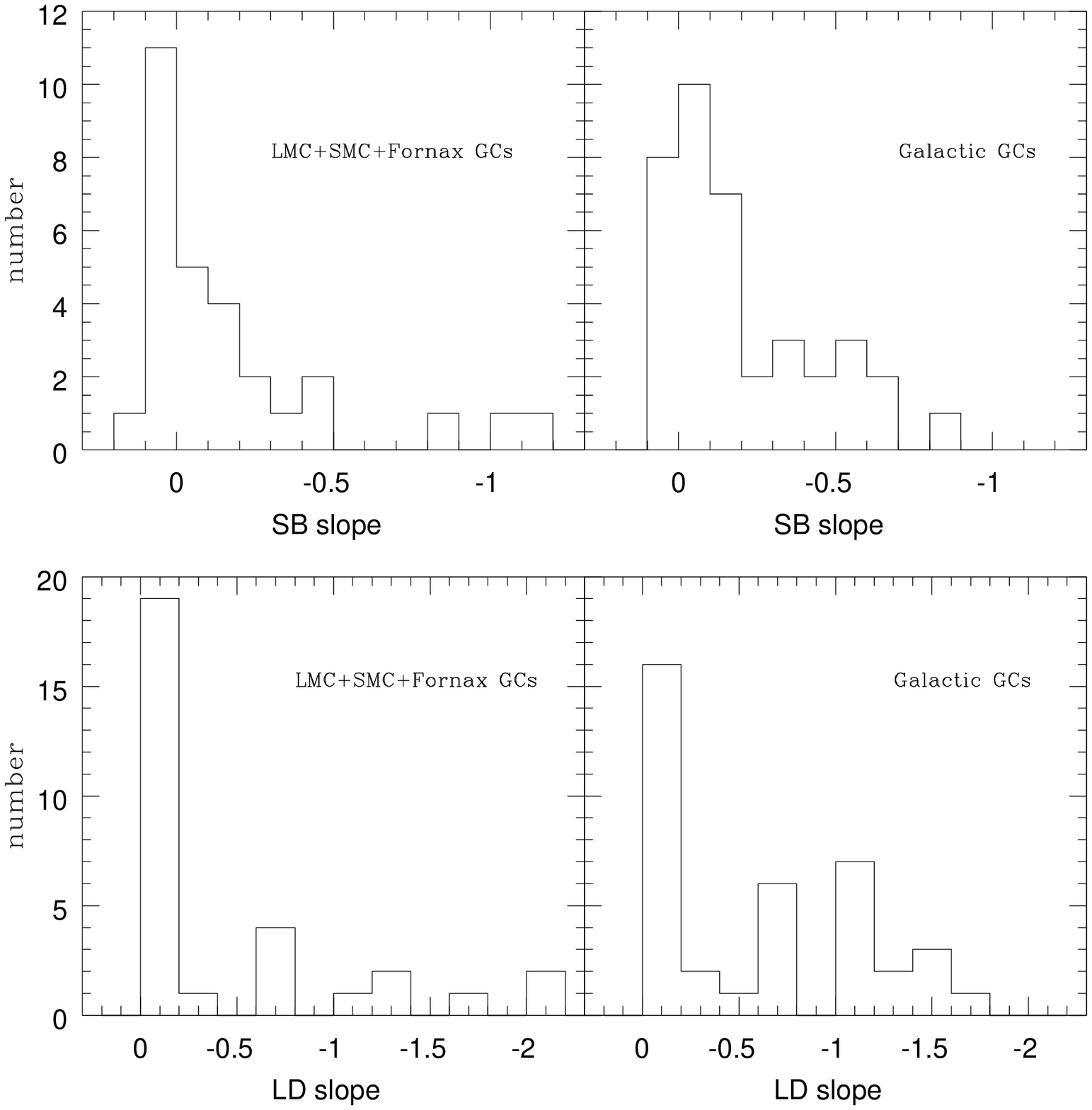,width=9.5cm,angle=0}}
\figcaption[Histograms of central logarithmic SB and LF
  slopes]{Histograms for surface brightness (top) and luminosity
  density (bottom) central logarithmic slopes for the LMC+SMC+Fornax
  sample (left panel) and the Galactic sample (right panel).}
\label{ch2f8}
\end{figure}

\begin{figure*}[t]
\centerline{\psfig{file=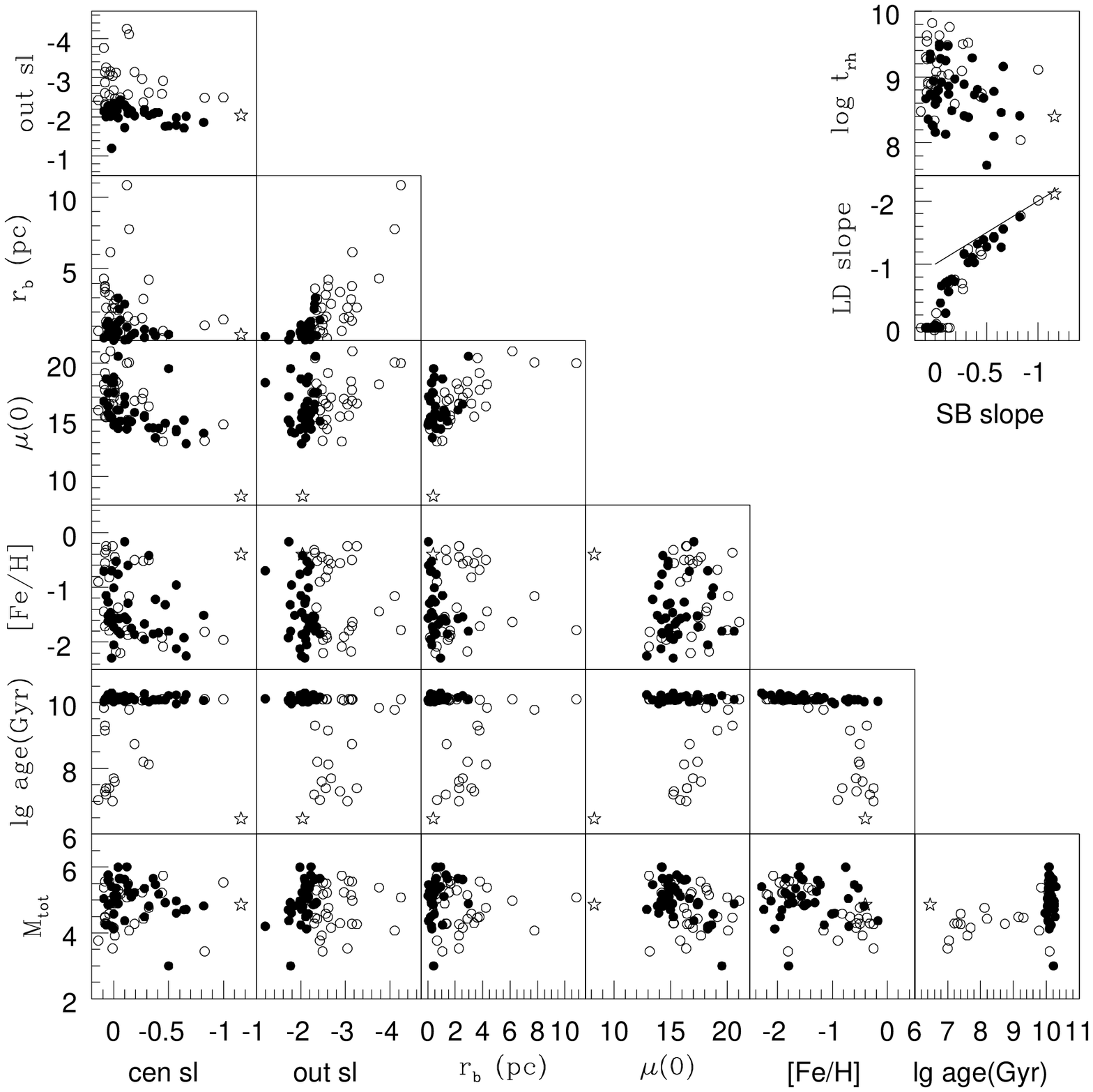,width=16.5cm,angle=0}}
\figcaption[Various physical quantities plotted against each
  other]{Surface brightness central and outer logarithmic slopes,
  logarithmic break radius (in parsecs), central surface brightness,
  metallicity, logarithmic age, and total mass plotted against each
  other for the LMC+SMC+Fornax sample (open points) and the Galactic
  sample (solid points). R136 is shown as a star symbol. We also show
  on the top right corner two panels with SB slope versus half-light
  relaxation time and versus LD slope (the solid line represents `LD
  slope = SB slope + 1'). The distances to the clusters are assumed to
  be 45 kpc for the LMC, 60 kpc for the SMC and 140 kpc for the Fornax
  dwarf galaxy.}
\label{ch2f9}
\end{figure*}

\subsection{Combining Two Samples}\label{two}

In order to explore possible correlations between physical quantities,
we combine the results for this sample with those for the Galactic
sample from paper~I. From now on, we will refer to the objects in the
LMC, SMC, and Fornax dwarf galaxy as the `satellite sample' or
`satellite clusters'. We compare the central slope measurements for
both samples by plotting the slope histograms side by side
(Fig~\ref{ch2f8}). We note that in both SB and LD central slopes, the
satellite sample extends to steeper slopes than the Galactic
sample. In total, 63\% of the satellite sample is consistent with
having flat cores, the remaining objects display a continuous
distribution of central slopes between 0 and $-1.4$ for surface
brightness, and between 0 and $-2.2$ for luminosity density. From
paper~I, we know that 50\% of the Galactic sample is consistent with
having flat cores, a smaller fraction than for the satellite
sample. For the Galactic sample we do not find any object with central
slopes steeper than $-0.8$ for SB or $-1.8$ for LD, we find two
objects (R136 and Fornax~3) steeper than that in the satellite
sample. Even when these differences are taken into account, the main
conclusion that the slope distributions are inconsistent with a
bimodal distribution of flat and PCC cores is the same for both
samples.

We plot a variety of physical quantities against each other in order
to explore for possible correlations in both samples. The metallicity
and age values are taken directly from MVM05. We observe in
Figure~\ref{ch2f9} that the younger clusters, which belong to the
satellite sample, have a narrower metallicity and total mass ranges
($-1<$[Fe/H]$<0$ and $3\times10^3M_\odot<M_{tot}<10^5M_\odot$
respectively) than the old ones. This can be due to the fact that for
our sample the young clusters sample a small linear age regime, so
they have fewer chances of populating the extreme mass regime. Both
metallicity and total mass do not show any clear correlations with
other physical quantities. The outer slope shows weak correlations for
the satellite sample in the sense that clusters with steeper outer
slopes seem to be older, have fainter central surface brightness, and
larger break radius. The Galactic clusters appear to have shallower
outer slopes than the satellite ones, but this could be due to an
effect of the difference between the Chebychev and the EFF fits used
for each sample (see section \ref{dsb}). The galactic clusters might
show steeper outer slopes if they were analyzed in the same way as the
satellite ones, or visceversa. We note that there seems to be a narrow
range of outer slopes between $-2$ and $-3$ for the clusters with
steep central slopes for both samples. There is a trend of clusters
with steeper central slopes having brighter central surface brightness
values. Every cluster with $\mu$(0)$<14.0$ mag/arcsec$^2$ has a
central logarithmic slope steeper than -0.4. Central surface
brightness seems to be fainter for older clusters, but this is only
observed for the satellite sample. Regarding the break radius, we
should clarify that the lack of clusters with break radii larger than
$\sim$4 pc in the Galactic sample is a selection effect due to the
fact that we required the core radius to fit on the WFPC field of
view. Since the satellite clusters are $4-14$ times further away than
the average Galactic cluster, we can include clusters with larger
break radius for the satellite sample. We note that all the clusters
with a central surface brightness brighter than $\sim$16
mag/arcsec$^2$ have break radii smaller than $\sim$2pc. Our measured
break radius follows the same trend observed for core radius versus
age by other authors \cite{els89,els92,deg02}. Clusters younger than
1~Gyr have break radii smaller than 4 pc, while older clusters span a
wide range of break radii. We notice that every cluster with central
SB slope steeper than $-0.5$ has a half-light relaxation time shorter
than 1 Gyr. Finally, the SB slope versus LD slope relation for the
satellite clusters lies right on top of the one observed for Galactic
clusters, which in turn is similar to the one observed for galaxies
\citep{geb96}.

\begin{deluxetable*}{lccrrrrccccrr}
\tablewidth{0pt}
\tabletypesize{\normalsize}
\tablecaption{\label{tbl3b}Results}
\tablehead{
\colhead{name} &
\colhead{$\mu_V(0)$}   &
\colhead{$r_b$} &
\colhead{$r_h$} &
\colhead{$lg~t_{rh}$} &
\colhead{$lg$ age} &
\colhead{log $M_{tot}$} &
\colhead{SB slope} &
\colhead{error} &
\colhead{LD slope} &\\
\colhead{} &
\colhead{(mag/arcsec$^2$)}   &
\colhead{arcsec} &
\colhead{arcsec} &
\colhead{years} &
\colhead{years} &
\colhead{$M_\odot$} &
\colhead{logarithmic} &
\colhead{} &
\colhead{logarithmic} &
}
\startdata
NGC~1466 & 18.4 & 13.2 & 24.3 & 9.21 & 10.10 & 5.15 & -0.02 & 0.18 & 0     \\
NGC~1651 & 20.5 & 16.6 & 71.2 & 9.63 & 9.30 & 4.47 & 0.08  & 0.20 & 0     \\
NGC~1711 & 17.0 & 11.7 & 30.7 & 8.96 & 7.70 & 4.16 &  0.00 & 0.18 & 0.00  \\
NGC~1754 & 17.4 & 5.6  & 15.2 & 8.81 & 10.11 & 4.90 & -0.01 & 0.12 & -0.23 \\
NGC~1786 & 15.0 & 7.6  & 14.9 & 9.04 & 10.11 & 5.50 & -0.13 & 0.14 & -0.70 \\
NGC~1805 & 16.4 & 10.4 & 17.0 & 8.33 & 7.00 & 3.53 &  0.01 & 0.13 & 0.04  \\
NGC~1818 & 16.4 & 10.5 & 26.9 & 8.92 & 7.40 & 4.27 &  0.07 & 0.14 & 0     \\
NGC~1835 & 15.4 & 7.5  & 11.4 & 8.89 & 10.11 & 5.58 & -0.04 & 0.16 & 0     \\
NGC~1866 & 16.2 & 19.4 & 49.7 & 9.52 & 8.12 & 4.76 & -0.32 & 0.12 & -1.24 \\
NGC~1868 & 16.7 & 6.3  & 16.2 & 8.59 & 8.74 & 4.28 & -0.19 & 0.13 & -0.76 \\
NGC~1898 & 18.2 & 9.7  & 42.0 & 9.62 & 10.11 & 5.29 & -0.04 & 0.13 & 0     \\
NGC~1916 & 13.1 & 3.0  &  8.2 & 8.75 & 10.11 & 5.74 & -0.45 & 0.16 & -1.15 \\
NGC~1984 & 15.8 & 3.1  & 18.4 & 8.48 & 7.06 & 3.77 &  0.14 & 0.08 & 0     \\
NGC~2004 & 15.3 & 5.5  & 21.0 & 8.77 & 7.30 & 4.30 &  0.08 & 0.16 & 0     \\
NGC~2005 & 14.2 & 0.7  & 10.6 & 8.80 & 10.11 & 5.22 & -0.44 & 0.13 & -1.21 \\
NGC~2019 & 13.2 & 4.9  & 11.3 & 8.04 & 10.11 & 3.44 & -0.83 & 0.15 & -1.77 \\
NGC~2031 & 17.4 & 13.3 & 59.6 & 9.50 & 8.20 & 4.43 & -0.20 & 0.11 & -0.61 \\
R136     &  8.3 & 1.9  &  8.4 & 8.40 & 6.48 & 4.86 & -1.16 & 0.21 & -2.11 \\
NGC~2100 & 15.2 & 6.0  & 25.6 & 8.89 & 7.20 & 4.29 &  0.07 & 0.14 & 0     \\
NGC~2214 & 17.7 & 10.4 & 42.6 & 9.08 & 7.60 & 3.92 & -0.01 & 0.13 & -0.06 \\
NGC~2257 & 21.0 & 28.3 & 68.7 & 9.82 & 10.11 & 4.98 &  0.03 & 0.16 & 0     \\
KRON~3   & 20.1 & 27.2 & 44.4 & 9.76 & 9.78 & 4.07 & -0.14 & 0.17 & 0     \\
NGC~121  & 17.6 & 13.3 & 23.2 & 9.54 & 10.08 & 5.56 &  0.08 & 0.14 & 0     \\
NGC~330  & 16.4 & 11.1 & 29.4 & 9.29 & 7.40 & 4.58 &  0.04 & 0.12 & 0     \\
NGC~411  & 19.1 & 13.2 & 30.1 & 0.27 & 9.15 & 4.49 &  0.08 & 0.25 & 0     \\
NGC~416  & 18.1 & 15.1 & 18.1 & 9.30 & 9.84 & 5.38 &  0.09 & 0.24 & 0     \\
FORNAX~2 & 20.0 & 16.0 & 12.4 & 9.48 & 10.11 & 5.08 & -0.12 & 0.13 & 0     \\
FORNAX~3 & 14.6 & 2.2  &  5.2 & 9.10 & 10.11 & 5.53 & -1.00 & 0.19 & -2.01 \\
FORNAX~4 & 16.8 & 2.3  &  6.1 & 9.09 & 10.06 & 5.24 & -0.26 & 0.13 & -0.70 \\
FORNAX~5 & 16.8 & 0.8  &  5.8 & 9.02 & 10.11 & 5.17 & -0.06 & 0.10 & 0.00  \\
\enddata
\end{deluxetable*}

\section{Summary and discussion}

We obtain central surface brightness profiles for 21 clusters in the
Large Magellanic Cloud, 5 in the Small Magellanic Cloud, and 4 in the
Fornax dwarf galaxy. We construct and analyze a large number of
simulated images in order find the most suitable way to obtain surface
brightness, as well as to estimate our uncertainties. The profiles are
constructed by measuring integrated light with a robust statistical
estimator. We combine $\it{HST}$/WFPC2 images in two filters (F555W
and F814W) when available and present profiles normalized to V-band
magnitudes.

When our results are compared with previous results that use different
analysis techniques, we find very good agreement for $\sim$60\% of the
sample. For the remaining 40\%, our central photometric points are
brighter than previous measurements. Most central surface brightness
values change from previously reported ones with values up to two
magnitudes brighter. For some objects in the sample, the new measured
surface brightness profile is no longer compatible with a flat core
parametric fit. The main reason for this difference is the increased
spatial resolution of $\it {HST}$, but also because we use a
non-parametric estimate as opposed to the traditional King model
fits. For some of the observed profiles the departures from a flat
core model are small, but significant. We confirm the existence of a
steep central cusp for three clusters previously classified as post
core-collapse. We also find a subpopulation of objects with shallow
cusps with logarithmic central slopes between $-0.2$ and $-0.5$. When
we plot a variety of physical quantities searching for correlations,
we find indications that the younger clusters tend to have smaller
break radius, shallower outer slopes, and brighter central surface
brightness. In particular, the youngest cluster in the sample, R136,
shows the steepest central profile and the brightest central surface
brightness. We also observe a clear correlation in which the clusters
with the steepest central slopes are the ones with the brightest
central surface brightness.

There have been two mechanisms explored for producing cusps in star
clusters: core-collapse and the presence of an intermediate mass black
hole in the center of the cluster. A detailed discussion and
references on this subject can be found on section~1.2 of paper~I. The
range of 3-dimensional density slopes is wider for core-collapse than
for black hole models, but they both center around the same value,
$\sim-1.65$. However, only the four clusters with the steepest
profiles in our sample fall in this range. In the case of
core-collapse, the slope depends on the mass of the stars used to
measure the profile and of those that dominate the mass of the core,
so this could extend the range toward shallower slopes. Another factor
of uncertainty is the time dependence of the core-collapse model when
the core goes through gravothermal oscillations. According to
Fokker-Planck simulations, a star cluster will spend a considerable
amount of time in between successive collapses, where the light
profile resembles a King model with a flat core. Unfortunately, these
models do not give enough details about the slope of the density
profile during intermediate stages of post-collapse bounce, or about
the time spent on intermediate stages.

As discussed in paper~I, an alternative way to explain the shallow
cusps, is by invoking the result by \citet{bau05}. They perform
simulations of stellar clusters with intermediate mass black holes in
their centers and find that, after a Hubble time, the projected
density distribution of the clusters show shallow cusps with slopes
around $-0.25$. For this sample, we find 4 objects that fall within
this regime, but only kinematical data can confirm the possible
existence of a central black hole for these objects.

The observed correlations with age observed for the Satellite sample
(section~\ref{two}) point out to the possibility of clusters having
very concentrated profiles during early stages of their evolution. In
particular the break radius-age relation observed here and by many
authors tells us that the size of cores depends on the dynamical
evolution of clusters. The input density profiles for various
dynamical simulations have almost always been characterized by King or
Plummer models. This could be biased toward large flat cores, when
more concentrated profiles could be more appropriated. This is true
for core-collapse models, as well as models containing a central black
hole.

Tables with the complete photometric points and smooth profiles for
every object in this sample can be found in the CDS-Vizier Service.

\acknowledgments

E. N. wants to thank Ralf Bender for his hospitality, as well as
Dougal Mackey and Dean McLaughlin for kindly and promptly sharing
their data. We acknowledge the grant under HST-AR-10315 awarded by the
Space Telescope Science Institute, which is operated by the
Association of the Universities for Research in Astronomy, Inc., for
NASA under contract NAS5-26555. We also acknowledge the technical
support from the Canadian Astronomy Data Centre, which is operated by
the Herzberg Institute of Astrophysics, National Research Council of
Canada. Finally, we acknowledge the support by CONACYT.


\begin{thebibliography}{}

\bibitem[Bastian \& Goodwin(2006)]{bas06}
{Bastian}, N., \& {Goodwin}, S.~P. 2006, MNRAS, 369, 9

\bibitem[\protect\citeauthoryear{{Bates} et~al.}{{Bates} et~al.}{1986}]{bat86}
{Bates}, D., {Lindstrom}, M., {Wahba}, G.,  \& {Yandell}, B. 1986, Technical
  Report 775, University of Wisconsin, Madison

\bibitem[\protect\citeauthoryear{{Baumgardt}, {Makino}, \& {Hut}}{{Baumgardt}
  et~al.}{2005}]{bau05}
{Baumgardt}, H., {Makino}, J.,  \& {Hut}, P. 2005, ApJ, 620, 238

\bibitem[\protect\citeauthoryear{{Beers}, {Flynn}, \& {Gebhardt}}{{Beers}
  et~al.}{1990}]{bee90}
{Beers}, T.~C., {Flynn}, K.,  \& {Gebhardt}, K. 1990, AJ, 100, 32

\bibitem[\protect\citeauthoryear{{Binney} \& {Tremaine}}{{Binney} \&
  {Tremaine}}{1987}]{bin87}
{Binney}, J.,  \& {Tremaine}, S. 1987, {Galactic dynamics} (Princeton, NJ,
  Princeton University Press, 1987, 747 p.)

\bibitem[\protect\citeauthoryear{{De Angeli} et~al.}{{De Angeli}
  et~al.}{2005}]{ang05}
{De Angeli}, F., {Piotto}, G., {Cassisi}, S., {Busso}, G., {Recio-Blanco}, A.,
  {Salaris}, M., {Aparicio}, A.,  \& {Rosenberg}, A. 2005, AJ, 130, 116

\bibitem[\protect\citeauthoryear{{de Grijs} et~al.}{{de Grijs}
  et~al.}{2002}]{deg02}
{de Grijs}, R., {Gilmore}, G.~F., {Mackey}, A.~D., {Wilkinson}, M.~I.,
  {Beaulieu}, S.~F., {Johnson}, R.~A.,  \& {Santiago}, B.~X. 2002, MNRAS, 337,
  597

\bibitem[\protect\citeauthoryear{{Elson}}{{Elson}}{1991}]{els91}
{Elson}, R.~A.~W. 1991, ApJS, 76, 185

\bibitem[\protect\citeauthoryear{{Elson}}{{Elson}}{1992}]{els92}
{Elson}, R.~A.~W. 1992, MNRAS, 256, 515

\bibitem[\protect\citeauthoryear{{Elson}, {Fall}, \& {Freeman}}{{Elson}
  et~al.}{1987}]{els87}
{Elson}, R.~A.~W., {Fall}, S.~M.,  \& {Freeman}, K.~C. 1987, ApJ, 323, 54

\bibitem[\protect\citeauthoryear{{Elson}, {Freeman}, \& {Lauer}}{{Elson}
  et~al.}{1989}]{els89}
{Elson}, R.~A.~W., {Freeman}, K.~C.,  \& {Lauer}, T.~R. 1989, ApJL, 347, L69

\bibitem[\protect\citeauthoryear{{Gebhardt} et~al.}{{Gebhardt}
  et~al.}{1996}]{geb96}
{Gebhardt}, K., et~al. 1996, AJ, 112, 105

\bibitem[Goodwin \& Bastian(2006)]{goo06} 
{Goodwin}, S.~P., \& {Bastian}, N.. 2006, MNRAS, 373, 752


\bibitem[\protect\citeauthoryear{{Jimenez} \& {Padoan}}{{Jimenez} \&
  {Padoan}}{1998}]{jim98}
{Jimenez}, R.,  \& {Padoan}, P. 1998, ApJ, 498, 704

\bibitem[\protect\citeauthoryear{{King}}{{King}}{1966}]{kin66}
{King}, I.~R. 1966, AJ, 71, 276

\bibitem[\protect\citeauthoryear{{Kontizas} \& {Kontizas}}{{Kontizas} \&
  {Kontizas}}{1983}]{kon83}
{Kontizas}, E.,  \& {Kontizas}, M. 1983, AAP, 52, 143

\bibitem[\protect\citeauthoryear{{Kontizas}, {Chrysovergis}, \&
  {Kontizas}}{{Kontizas} et~al.}{1987}]{kon87a}
{Kontizas}, M., {Chrysovergis}, M.,  \& {Kontizas}, E. 1987, AAP, 68, 147

\bibitem[\protect\citeauthoryear{{Kontizas}, {Theodossiou}, \&
  {Kontizas}}{{Kontizas} et~al.}{1986}]{kon86}
{Kontizas}, M., {Theodossiou}, E.,  \& {Kontizas}, E. 1986, AAP, 65, 207

\bibitem[\protect\citeauthoryear{{Lauer} et~al.}{{Lauer} et~al.}{1995}]{lau95}
{Lauer}, T.~R., et~al. 1995, AJ, 110, 2622

\bibitem[\protect\citeauthoryear{{Mackey} \& {Gilmore}}{{Mackey} \&
  {Gilmore}}{2003b}]{mac03a}
{Mackey}, A.~D.,  \& {Gilmore}, G.~F. 2003b, MNRAS, 338, 85

\bibitem[\protect\citeauthoryear{{Mackey} \& {Gilmore}}{{Mackey} \&
  {Gilmore}}{2003a}]{mac03b}
{Mackey}, A.~D.,  \& {Gilmore}, G.~F. 2003a, MNRAS, 338, 120

\bibitem[\protect\citeauthoryear{{Mackey} \& {Gilmore}}{{Mackey} \&
  {Gilmore}}{2003c}]{mac03c}
{Mackey}, A.~D.,  \& {Gilmore}, G.~F. 2003c, MNRAS, 340, 175

\bibitem[\protect\citeauthoryear{{Massey} \& {Hunter}}{{Massey} \&
  {Hunter}}{1998}]{mas98}
{Massey}, P.,  \& {Hunter}, D.~A. 1998, \apj, 493, 180

\bibitem[\protect\citeauthoryear{{Mateo}}{{Mateo}}{1987}]{mat87}
{Mateo}, M. 1987, ApJL, 323, L41

\bibitem[\protect\citeauthoryear{{McLaughlin} \& {van der Marel}}{{McLaughlin}
  \& {van der Marel}}{2005}]{mcl05}
{McLaughlin}, D.~E.,  \& {van der Marel}, R.~P. 2005, ApJS, 161, 304

\bibitem[\protect\citeauthoryear{{Noyola} \& {Gebhardt}}{{Noyola} \&
  {Gebhardt}}{2006}]{noy06}
{Noyola}, E.,  \& {Gebhardt}, K. 2006, \aj, 132, 447

\bibitem[\protect\citeauthoryear{{Rodgers} \& {Roberts}}{{Rodgers} \&
  {Roberts}}{1994}]{rod94}
{Rodgers}, A.~W.,  \& {Roberts}, W.~H. 1994, AJ, 107, 1737

\bibitem[\protect\citeauthoryear{{Salaris} \& {Weiss}}{{Salaris} \&
  {Weiss}}{2002}]{sal02}
{Salaris}, M.,  \& {Weiss}, A. 2002, AAP, 388, 492

\bibitem[\protect\citeauthoryear{{Smith} et~al.}{{Smith} et~al.}{1996}]{smi96}
{Smith}, E.~O., {Neill}, J.~D., {Mighell}, K.~J.,  \& {Rich}, R.~M. 1996, AJ,
  111, 1596

\bibitem[\protect\citeauthoryear{{Stetson}}{{Stetson}}{1987}]{ste87}
{Stetson}, P.~B. 1987, PASP, 99, 191

\bibitem[\protect\citeauthoryear{{Trager}, {King}, \& {Djorgovski}}{{Trager}
  et~al.}{1995}]{tra95}
{Trager}, S.~C., {King}, I.~R.,  \& {Djorgovski}, S. 1995, AJ, 109, 218

\bibitem[\protect\citeauthoryear{Wahba}{Wahba}{1980}]{wab80}
Wahba, G. 1980, Technical Report 595, University of Wisconsin, Madison

\bibitem[\protect\citeauthoryear{Wahba \& Wang}{Wahba \& Wang}{1990}]{wah90}
Wahba, G.,  \& Wang, Y. 1990, Communications in Statistics, Part A -- Theory
  and Methods, 19, 1685

\bibitem[\protect\citeauthoryear{{Wilson}}{{Wilson}}{1975}]{wil75}
{Wilson}, C.~P. 1975, AJ, 80, 175

\end{thebibliography}
\end{document}